\documentclass{vgtc}                          

\ifpdf
  \pdfoutput=1\relax                   
  \usepackage{graphicx}                
  \DeclareGraphicsExtensions{.pdf,.png,.jpg,.jpeg} 
\else
  \usepackage{graphicx}                
  \DeclareGraphicsExtensions{.eps}     
\fi%

\graphicspath{{figures/}{pictures/}{images/}{./}} 

\usepackage{microtype}       
\usepackage{amsmath}
\PassOptionsToPackage{warn}{textcomp}  
\usepackage{textcomp}                  
\usepackage{mathptmx}                  
\usepackage{times}                     
\usepackage{cite}                      
\usepackage{tabu}                      
\usepackage{booktabs}                  

\onlineid{0}

\vgtccategory{Research}

\vgtcinsertpkg

\title{SUMART: SUMmARizing Translation from Wordy to Concise Expression}

\author{Naoto Nishida\\
        \scriptsize The University of Tokyo \\ %
        \scriptsize nawta@g.ecc.u-tokyo.ac.jp
\and Jun Rekimoto\\
     \parbox{1.4in}{\scriptsize \centering The University of Tokyo \\ Sony CSL Kyoto \\ rekimoto@acm.org}}

\teaser{
  \centering
  \includegraphics[width=\linewidth]{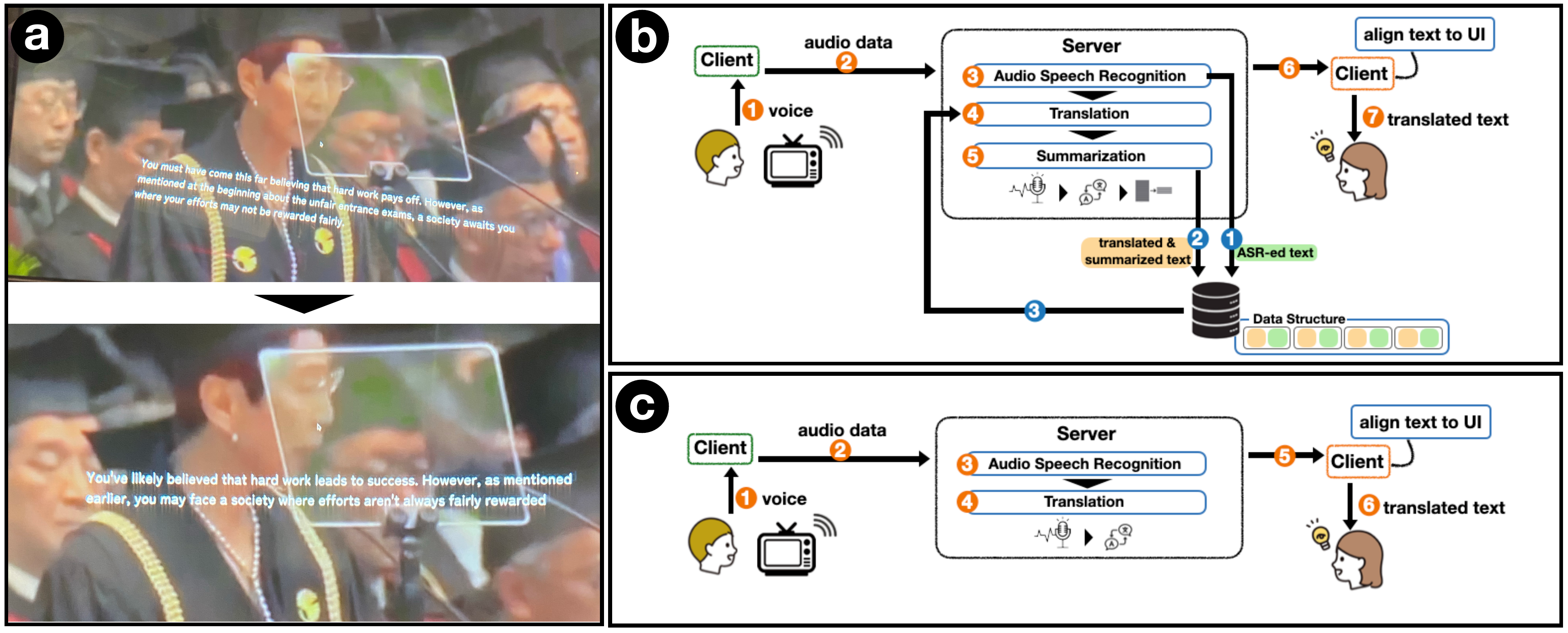}
  \caption{SUMART is a method for summarizing and compressing the volume of verbose subtitle translations to enable users to quickly comprehend what the speaker says. It is intended for users who want a big-picture and fast understanding of the conversation, audio, video content, and speech in a foreign language. a)~SUMART alters the wordy sentences to concise expressions. b)~Training phase of SUMART. SUMART collects the pair of ASR results and the translated\&summarized results as the training data, enabling people to use the translated caption system. c)~In practical uses, SUMART uses a translation model fine-tuned by the paired data collected in the training phase, resulting in a more concise translation output.}
  \label{fig: teaser}
}

\abstract{
We propose SUMART, a method for summarizing and compressing the volume of verbose subtitle translations. 
SUMART is designed for understanding translated captions (e.g., interlingual conversations via subtitle translation or when watching movies in foreign language audio and translated captions).
SUMART is intended for users who want a big-picture and fast understanding of the conversation, audio, video content, and speech in a foreign language.
During the training data collection, when a speaker makes a verbose statement, SUMART employs a large language model on-site to compress the volume of subtitles. This compressed data is then stored in a database for fine-tuning purposes. 
Later, SUMART uses data pairs from those non-compressed ASR results and compressed translated results for fine-tuning the translation model to generate more concise translations for practical uses. 
In practical applications, SUMART utilizes this trained model to produce concise translation results.
Furthermore, as a practical application, we developed an application that allows conversations using subtitle translation in augmented reality spaces. 
As a pilot study, we conducted qualitative surveys using a SUMART prototype and a survey on the summarization model for SUMART. 
We envision the most effective use case of this system is where users need to consume a lot of information quickly (e.g., Speech, lectures, podcasts, Q\&A in conferences).
} 

\CCScatlist{
  \CCScatTwelve{Human-centered computing}{Interaction design}{Activity centered design}{};
  \CCScatTwelve{Human-centered computing}{Visu\-al\-iza\-tion}{Visualization design and evaluation methods}{}
}

\begin{document}



\maketitle

\section{Introduction} 
In recent years, there has been an increase in the use of subtitle translation for communication. For instance, Japanese cast members at maid cafes converse with foreigners using PokeTalk in Akihabara. 
The iPhone's built-in translation app and Google Translate also feature a conversation mode that facilitates interactive communication. 
Moreover, companies like Google and TCL have released demo videos and exhibited prototypes displaying subtitles on smart glasses for real-time translation. 

However, there has been a lack of field research on the practical use of these real-time subtitle translation systems. 
Our pilot study found that using these subtitle translation apps for conversation can lead to a high cognitive load and increased processing time due to the volume of subtitles. 
Therefore, we propose SUMART, which summarizes verbose subtitles and incorporates a human-in-the-loop approach for data collection to improve the model (Figure ~\ref{fig: teaser} a).
SUMART is intended for users who want a big-picture and fast understanding of the conversation, audio, video content, and speech in a foreign language.
Thereby, we envision the most effective use case of this system is where users need to consume a lot of information quickly (e.g., speech, lectures, podcasts, scientific Q\&A in conferences).
We employed translated captions as modality because it is expected to be employed on emerging state-of-the-art smartglasses translation function.

Furthermore, as an application example, we have developed an application that projects subtitles onto AR glasses. We have also conducted a preliminary study on the summarization model used in SUMART.

\section{How to Convey Information Time-Efficiently}
Here, we discuss the temporal lag in conversations between speakers of different languages using translated subtitles~\cite{stuart}. 
The typical process during a conversation between Speaker A and Speaker B involves several steps: 1) Speaker A reads the translated subtitles, 2) Speaker A processes the information to formulate a response, 3) Speaker A speaks their response, 4) Speaker A's response is translated, 5) Speaker B reads the translated subtitles, 6) Speaker B processes the information to formulate a response, 7) Speaker B speaks their response, and 8) Speaker B's response is translated...
These steps are generally processed in a waterfall manner, although some may occur concurrently, such as thinking while speaking.

Considering the speed of each part, it is noted that the average reading rate for adult native English speakers is about 238 words per minute (wpm), and for children, it ranges from 53 to 204 wpm \cite{BRYSBAERT2019104047, Hasbrouck2017-em}. 
Meanwhile, the speaking rate for various native speakers of English is approximately 120-160 wpm \cite{Yuan2006-we}, indicating that the average amount of information conveyed in speech, in terms of English word count, is around 150 wpm regardless of language.

To exchange more information per unit of time in conversations using translated subtitles, we propose two approaches: 1) Increase the amount of information in the conversation, and 2) Improve the tempo of the conversation. 
For the first approach, methods like VisualCaptions and RealityTalk enhance the information content in conversations by supplementing spoken content with related images or AR effects \cite{Liao:2022:10.1145/3526113.3545702, Liu:2023:10.1145/3544548.3581566}. 
The second approach involves speeding up AI processes like translation; an area actively pursued in AI research. However, attempts to expedite human processes such as reading and speaking are lacking. 

Therefore, we propose a system designed to accelerate the reading process to quicken the human side of these conversations.

\section{SUMART}

We present SUMART, a system incorporating summarization tasks and a human-in-the-loop mechanism into traditional translation tasks, creating a model that outputs concise translations. 

In the following sections, we will address the temporal bottlenecks encountered in conversations using subtitle translations and delve into the theoretical underpinnings of SUMART. 
This will be accompanied by a detailed description of the prototype we have developed and the feedback received regarding this prototype.


\subsection{Base Theory}
Here, we examine the efficiency of subtitle translation and summarization in conversation. 
First, we denote the number of words a speaker speaks as {\it wc}. 
Since humans speak at an approximate rate of 150 words per minute (wpm) in any language~\cite{Yuan2006-we}, the time taken for a speaker to make a statement is $\frac{1}{150} * wc$ minutes. 
If we define the compression rate of summarization as $\sigma (0 < \sigma \le 1)$, the time taken for cognitive processing in seconds as $\gamma$, the time for translation as $t\_{trans}$ in seconds, and the time for summarization as $t_{sum}$, then the total time in seconds for steps 1 to 4 of the previous section's process can be expressed by the following formula:

\begin{equation}
\begin{split}
    &(\frac{1}{238} * wc * \sigma + \frac{1}{150} * wc) * 60 + \gamma + t_{trans} + t_{sum} [sec]\\ 
    &= \epsilon * wc + \gamma + t_{trans} + t_{sum} [sec]
\end{split}
\end{equation}

Here, $\epsilon$, which falls in the range $60 * \frac{1}{150} < \epsilon \le 60 * (\frac{1}{150} + \frac{1}{238})$, represents the seconds per word. 
Considering that the average English speech length per utterance is about 20 words, the maximum reduction in transmission time per statement can be calculated as $20 * (\epsilon_{max} - \epsilon_{min}) \simeq 20 * ((\frac{60}{238} + \frac{60}{150}) - \frac{60}{150}) \simeq 5.04...$ seconds. This implies a potential maximum reduction of approximately 5 seconds in transmission time per utterance. 
Of course, a compression rate $\sigma$ of 0 is unrealistic as it would imply telepathic communication.
However, even with a compression rate of $\frac{2}{3}$, this can lead to a reduction of about $5 * \frac{1}{3} \simeq 1.66...$ seconds. Considering the duration of conversations over a long period, this can significantly improve the efficiency of information transmission. 
Moreover, concerning the time taken for the summarization task $t_{sum}$, SUMART is designed to learn to produce shorter translations using post-summarization translations and pre-translation text, making $t_{sum}$ effectively zero in practical applications.

\subsection{System Architecture}
\subsubsection{Data Collection and Practical Use}
We describe the workflow for collecting training data for SUMART, as illustrated in Fig. \ref{fig: teaser}~b. The process involves two primary components, represented in orange and blue in the architecture diagram.

The orange section represents the architecture for information transmission from speakers or TVs who speak Language A to those who understand Language B, using subtitle translation. The steps are as follows: 1) The speaker's or TV's voice is inputted into the client computer. 2) This audio data is then sent to the server. 3) On the server, Audio Speech Recognition is performed. 4) The speech is then translated. 5) This is followed by summarization. 6) and 7) Finally, the summarized translation is sent to the listener.

On the other hand, the blue section of the workflow is dedicated to collecting data for training subsequent translation models. This involves: 1) Storing the speech recognition text of the speaker's utterances and 2) the translated and summarized version of these utterances as paired data in a database. This data is then used to fine-tune the translation model as required.

In practical application, the fine-tuned model is utilized in the same workflow as the orange section of the Training Phase (Fig. \ref{fig: teaser} c). This means that the information transmission process from speakers or TVs who speak Language A to those who understand Language B is conducted using subtitle translation, following the steps laid out in the training phase.

\begin{table}[]
\caption{Results of abstractive summarization task of each model.}
\label{tab:my-table}
\centering
\begin{tabular}{lrr}
\hline
Model & \begin{tabular}[c]{@{}l@{}}Average required time \\ for execution (SD)\end{tabular} & \begin{tabular}[c]{@{}l@{}}The num of \\ output token\end{tabular} \\ \hline
Azure API         & 6.68 (0.612)   & 39.0 \\
PEGASU\_LARGE & 0.318 (0.0181) & 13.0 \\
T5\_LARGE      & 0.332 (0.341)  & 27.0 \\
chatGPT       & 2.45 (0.350)   & 30.4 \\ \hline
\end{tabular}
\end{table}

\begin{figure}[tb]
 \centering 
 \includegraphics[width=\columnwidth]{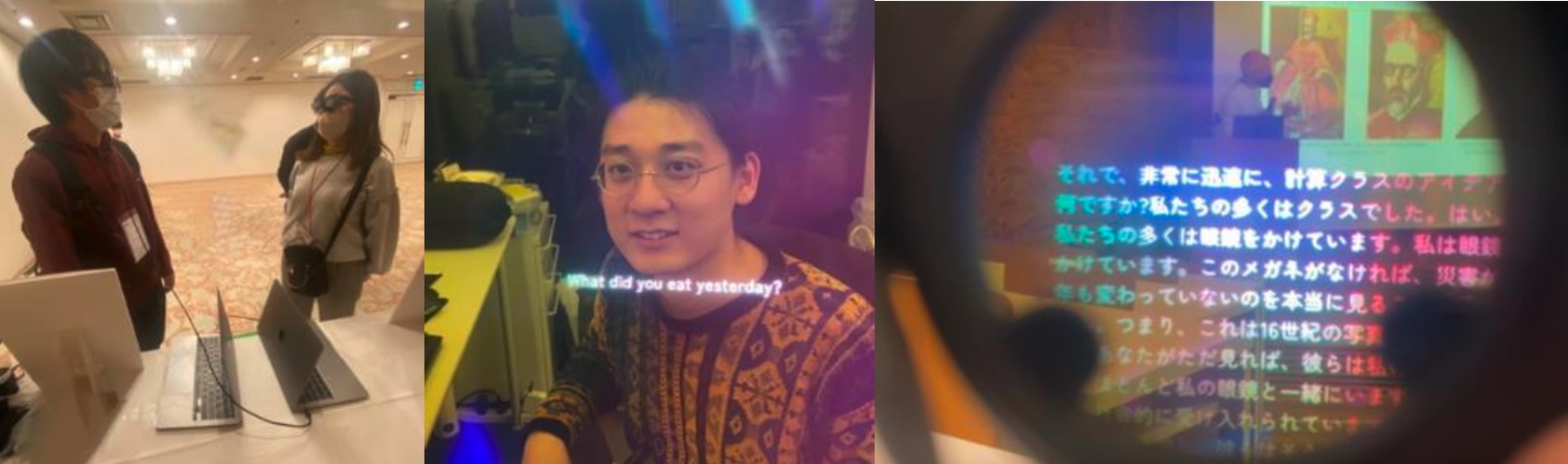}
 \caption{Examples of people using SUMART and user interface of SUMART prototype.}
 \label{fig:usecase}
\end{figure}

\subsubsection{Prototype}
We created a prototype to investigate how people converse using subtitle translations by displaying voice recognition results as translated subtitles on AR glasses. For hardware, we utilized Nreal Light. On the software side, we developed a client using Unity and C\# and a server using Python. Additionally, we employed Microsoft Azure API on the server side. The server and client communication was facilitated using socket communication with ZMQ. The system's architecture is illustrated in Fig \ref{fig: teaser} b and c. Users conducted conversations using AR glasses connected to a PC. The nature of these conversations and the user's point of view are depicted in Fig. \ref{fig:usecase}.

Regarding the summarization model, we conducted a survey, the results of which are presented in Table \ref{tab:my-table}. We input the sentence of the second paragraph from a web article~\cite{bushhospitalized}. 
Based on this survey, the ChatGPT API was selected due to its minimal latency in summarization tasks.
We used a simple prompt as "Summarize this sentence: \{user input\}" in case of using chatGPT.

\subsection{User Feedback, Discussion and Future Work}
We received feedback from users who tested our prototype. The feedback encompassed various aspects: audio-related issues such as source separation, natural language processing concerns like translation accuracy and mechanisms for ensuring it, and user interface elements like subtitle display positioning. 

Additionally, although we used simple prompts for summarization, there was also a suggestion that adhering to the Prompt Principle~\cite{bsharat2023principled}, it might be feasible to incorporate more of the conversational context or consider the expertise of the conversation partner for tailored subtitle adjustments. 

Besides, we found that most of people felt the summarization does not need in the case of daily conversation scenarios, since they speak easy and short sentences usually and sometimes they intentionally make verbose statements (e.g., euphemism, jokes or humors). 
However, they strongly felt the needs of the summarization when the speaker themselves start to speak gibberish or too fast and long, leading to our confirmation that this system's most effective scenarios are when translating long and fast paragraphs such as speech, news, podcasts, lectures, scientific questions, and so on.

The participants also suggested to integrate the contexts to the output, such as the situations or other paralingual features. For example, one participant mentioned that reflecting the others' plausible emotions from their facial expressions, voice tones, and way of saying could lead to precise translation results. This function is also useful, considering past works successfully integrated them into the change of translation results~\cite{Ahmed:2013:10.1145/2451176.2451197, Szekely:2014:10.1007/s12193-013-0128-x}.

As part of our future work, we plan to use these suggestions to create a design space and develop an improved version of SUMART.

\acknowledgments{
This project is supported by JST Moonshot R\&D Grant JPMJMS2012, JST CREST Grant JPMJCR17A3, the commissioned research by NICT Japan Grant JPJ012368C02901, and JST ASPIRE Program Grant Number JPMJAP2327.
}

\bibliographystyle{abbrv-doi}

\bibliography{template}
\end{document}